\documentclass[reprint,a4paper,prb,fleqn,showpacs,citeautoscript,superscriptaddress]{revtex4-1}

\usepackage{amssymb}
\usepackage{amsmath}
\usepackage{graphicx}
\usepackage{textcomp}

\usepackage[varg]{txfonts}   % Times typeface, used for some special symbols

\usepackage{dcolumn}

\usepackage[colorlinks=false,
            bookmarks=true,
            bookmarksnumbered=true]{hyperref}
\hypersetup{pdfborder = {0 0 0}}

\newcommand{\etal}{\textit{et al.}}
\newcommand{\half}{\ensuremath{\tfrac{1}{2}}}
\newcommand{\degree}{\mbox{$^\circ$}}
\newcommand{\degreeC}{\mbox{$^\circ$C}}
\newcommand{\BimTen}{\mbox{Bi$_m$Te$_n$}}
\begin{document}

\preprint{\today}

\title{Atomic Ordering in Cubic Bismuth Telluride Alloy Phases at High Pressure}

\author{I. Loa}
\email[Corresponding author:~E-mail~]{I.Loa@ed.ac.uk}
\affiliation{SUPA, School of Physics and Astronomy, and Centre for Science
at Extreme Conditions, The University of Edinburgh, Edinburgh, EH9 3FD, United Kingdom}

\author{J.-W. G. Bos}
\author{R. A. Downie}
\affiliation{Institute of Chemical Sciences and Centre for Advanced Energy Storage and Recovery, Heriot-Watt University,
Edinburgh, EH14 4AS, United Kingdom}

\author{K. Syassen}
\affiliation{Max-Planck-Institut f\"{u}r Festk\"{o}rperforschung, Heisenbergstr.\ 1,
70569 Stuttgart, Germany}

\date{\today}

\begin{abstract}
 Pressure-induced transitions from ordered intermetallic phases to substitutional alloys to
 semi-ordered phases were studied in a series of bismuth tellurides. Using angle-dispersive
 x-ray diffraction, the compounds Bi$_4$Te$_5$, BiTe, and Bi$_2$Te were observed to form
 alloys with the disordered body-centered cubic (bcc) crystal structure upon compression to
 above 14--19~GPa at room temperature. The BiTe and Bi$_2$Te alloys and the previously
 discovered high-pressure alloys of Bi$_2$Te$_3$ and Bi$_4$Te$_3$ were all found to show atomic
 ordering after gentle annealing at very moderate temperatures of $\sim$100{\mbox{\textdegree}}C. Upon annealing,
 BiTe transforms from the bcc to the B2 (CsCl) crystal structure type, and the other phases
 adopt semi-disordered variants thereof, featuring substitutional disorder on one of the two
 crystallographic sites. The transition pressures and atomic volumes of the alloy phases show
 systematic variations across the \BimTen\ series including the end members Bi and Te.
 First-principles calculations were performed to characterize the electronic structure and
 chemical bonding properties of B2-type BiTe and to identify the driving forces of the
 ordering transition. The calculated Fermi surface of B2-type BiTe has an intricate structure
 and is predicted to undergo three topological changes between 20 and 60~GPa.
\end{abstract}


\maketitle

%----------------------------------------------------------------------------
% Introduction
%----------------------------------------------------------------------------

\section{Introduction}
Bismuth tellurides, \BimTen, are a class of intermetallic compounds with intriguing physical
properties. Bi$_2$Te$_3$ is one of the best bulk thermoelectric materials for operation near
room temperature\cite{Gol14}; it becomes superconducting under
pressure\cite{II72,ETNO10,ZZWZ11,ZSCZ11,MTZU14}; it has been reported to undergo a
pressure-induced  electronic topological (Lifshitz)
transition\cite{IKK97,JKCS07,ETNO10,PGST11,ZLZK13}, and it has recently attracted much
interest in the context of topological insulators\cite{HK10,QZ11,CJFG13}. Like Bi$_2$Te$_3$,
the compound Bi$_4$Te$_3$ also becomes superconducting under pressure\cite{JLSS11} after
turning from a narrow-bandgap semiconductor into a metal. The structural changes in
Bi$_2$Te$_3$ and Bi$_4$Te$_3$ under compression have been studied previously, and both
intermetallic compounds were observed to transform to alloys with a disordered
body-centered-cubic (bcc) crystal structure above $\sim$15~GPa at room
temperature\cite{EONI11,ZWWL11,JLSS11}. The two materials are members of the infinitely
adaptive series of compounds with compositions $\mathrm{(Bi_{2})_\mu \cdot\,(Bi_2Te_3)_\nu}$,
and their crystal structures at ambient conditions are all superlattice structures consisting
of hexagonal Bi$_{2}$ and Bi$_2$Te$_3$ blocks \cite{BZLO07,BFDM12a,CJFG13}. The materials with
compositions near $\mu$:$\nu$ = 2:1 (e.g.\ Bi$_2$Te) are promising $p$-type thermoelectrics
\cite{BZLO07,BFDM12a}.

Here we show that the existence of high-pressure bcc alloy phases is a common feature across
the \BimTen\ series, and that these phases are thermally rather unstable --- they have a
propensity for atomic ordering upon gentle annealing at only $\sim$100{\mbox{\textdegree}}C under pressure.
During annealing, the BiTe compound adopts the fully-ordered B2 (CsCl-type) structure, and the
other compounds semi-disordered variants thereof. First-principles electronic structure
calculations are used to investigate the chemical bonding and electronic structure of B2-type
BiTe and to gain insight on the driving forces for the ordering transition.

\section{Methods}

The compounds Bi$_4$Te$_5$, BiTe, Bi$_4$Te$_3$, and Bi$_2$Te were synthesized as
polycrystalline pellets by solid-state reaction from the elements as described previously
\cite{BZLO07,BFDM12a}. A single crystal of Bi$_2$Te$_3$ was provided by D. L. Sun and C. T.
Lin (Max-Planck-Institut f\"{u}r Festk\"{o}rperforschung, Stuttgart). Each material was manually
ground to a fine powder and loaded into diamond anvil cells (DACs) for high-pressure
generation. In the majority of the experiments, condensed helium was used as the pressure
transmitting medium; some additional experiments that focussed on the effect of annealing the
samples at $\sim$100\degreeC\ were performed with nitrogen or a 4:1 methanol-ethanol mixture
as the pressure transmitting medium. Pressures where determined with the ruby fluorescence
method using the calibration of Ref.~\onlinecite{MXB86}.

Monochromatic powder x-ray diffraction patterns were measured in the Debye-Scherrer geometry
on beamline ID09a of the European Synchrotron Radiation Facility (ESRF) and beamline I15 of
Diamond Light Source (DLS), UK. Monochromatic x-rays of wavelength $\lambda \approx 0.415$~{\AA}
were focused to a spot size of $\sim$20~{\mbox{\textmu}}m at the sample, and the two-dimensional diffraction
images were recorded with a mar555 and a Perkin Elmer XRD1621 flat panel detector at the ESRF
and DLS, respectively. The DACs were rotated by ${\pm}3\mbox{--}5$\degree\ during the exposure to
improve the powder averaging. The diffraction images were integrated azimuthally with the
Fit2D software\cite{soft:Fit2d} to yield conventional intensity vs $2\Theta$ diffraction
diagrams, and then analyzed with the Rietveld method using the program Jana2006\cite{PDP14}.

Electronic structure calculations of BiTe were performed in the framework of density
functional theory (DFT) and the full-potential augmented-plane-wave + local orbital (APW+lo)
approach as implemented in the \textsc{Wien2k} \cite{soft:Wien2k} and
\textsc{Elk}\cite{soft:Elk2.3.16} codes. The $5d$, $6s$, and $6p$ orbitals of Bi and the $4d$,
$5s$, and $5p$ orbitals of Te were treated as valence states. Exchange and correlation effects
were treated with the revised generalized gradient approximation for solids (PBEsol)
\cite{PRCV08}. Spin-orbit coupling was previously found to have a significant
effect\cite{MSJ97,TRDM92} on the electronic structure of Bi$_2$Te$_3$ and was therefore
included in the present calculations. Regular Monkhorst-Pack grids of k-points, typically
$20\times20\times20$, were used for the Brillouin-zone sampling.\cite{note:DFT-full-details}
The electronic band structure and density of states obtained with the two codes were in
complete agreement. To quantify the charge transfer between Bi and Te in BiTe, the
\textsc{critic} code\cite{OBPL09} was used to perform Bader analyses of the charge density
distribution obtained with Wien2k. The electron localization function (ELF) \cite{BE90,SNWF97}
was calculated using the ELK code.

\section{Results and Discussion}
\subsection{Disordered bcc Phases}

\begin{figure}[t]
     \centering
     \includegraphics{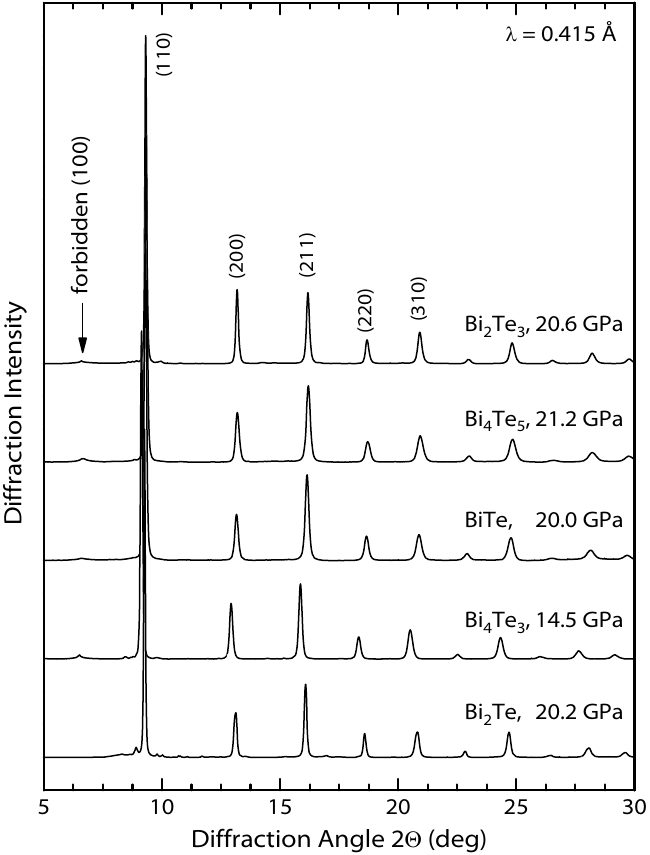}
     \caption{Powder x-ray diffraction patterns of the disordered-bcc phases of \BimTen\
              after compression at room temperature. All phases except Bi$_2$Te show a
              weak feature at the position of the (100) reflection, which is absent in an
              ideal bcc phase. Note that the reflection intensities of Bi$_2$Te$_3$ have
              been affected by preferred orientation of the powder sample, and the very
              weak features near 9\degree\ originate from the nitrogen pressure medium or
              the remainder of a lower-pressure phase. }
     \label{fig:Bi-Te_bcc_patterns}
\end{figure}

All five compounds were compressed at room temperature, and after one or more phase
transitions, each of them was found to transform to a phase with the diffraction pattern
characteristic of a bcc crystal, as illustrated in Figure~\ref{fig:Bi-Te_bcc_patterns}.
This suggests that disordered bcc alloy phases at high pressure are a general feature of the
compounds across the Bi-Te system. With the exception of the Bi$_2$Te pattern, the diffraction
patterns show a very weak feature at the position of the (100) reflection, which is forbidden
for a monatomic or ideal bcc phase. In the alloy phases, the presence of this reflection
indicates a slight deviation from a completely random distribution of the Bi and Te atoms,
i.e., the occurrence of some short-range correlations. Overall, however, Rietveld analysis of
the diffraction patterns confirms them to be consistent with the model of disordered Bi-Te
alloys as reported previously for Bi$_2$Te$_3$ and Bi$_4$Te$_3$
(Refs.~\onlinecite{EONI11,JLSS11}).

\begin{figure}[bt]
     \centering
     \includegraphics{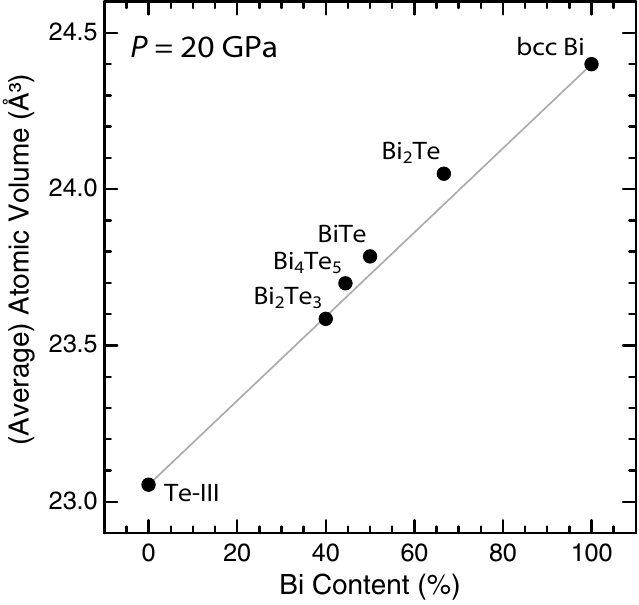}
     \caption{Average atomic volumes of the bcc alloy phases of \BimTen\ and atomic
              volumes of Bi and Te at 20 GPa.\cite{AKS02,HM03} Note that the crystal
              structure of Te at 20 GPa is incommensurately-modulated body-centered
              monoclinic (Te-III)\cite{HM03}. The straight line indicates a linear
              variation of the volume with composition according to Vegard{'}s law.  }
     \label{fig:Bi-Te_bcc_Volumes}
\end{figure}

Figure~\ref{fig:Bi-Te_bcc_Volumes} shows the average atomic volumes for the alloy phases at
20~GPa as a function of composition. The atomic volumes of Bi and Te at the same pressure are
shown for comparison. Note that Bi adopts the bcc structure at this pressure, whereas Te has
an incommensurately-modulated body-centered monoclinic structure\cite{HM03} and transforms to
bcc only at $\sim$29~GPa. There is a continuous evolution of the atomic volume across the
series with only a slight deviation from Vegard{'}s law. Bi$_2$Te shows the most notable
deviation (0.4\%). Einaga \etal\cite{EONI11} performed a similar analysis for Bi$_2$Te$_3$
between 23 and 30~GPa and concluded that there is a substantial deviation from Vegard{'}s law,
which was attributed to {``}remaining strong ionic-covalent bonds{''}. The important difference
between the two analyses is that Einaga \etal\ used an extrapolation of the equation of state
of bcc Te\cite{PH88} to lower pressures whereas we used the actual atomic volume\cite{HM03} of
Te at 20~GPa as a reference. The former has the advantage of referring to the same crystal
structure for the whole series of compositions, whereas the latter uses only the
experimentally observed volumes, irrespective of the crystal structure. We should note that
the deviation from Vegard{'}s law in Einaga{'}s analysis becomes significantly smaller if, instead
of the of bcc Te equation-of-state data by Parthasarathy and Holzapfel\cite{PH88}, one uses
the more recent data by Hejny and McMahon\cite{HM03}. Overall, we see little evidence for a
significant deviation from Vegard{'}s law. The nature of the bonding in the Bi-Te alloys at high
pressure, on the other hand, remains a pertinent question, to which we will return later.

\begin{figure}[bt]
     \centering
     \includegraphics{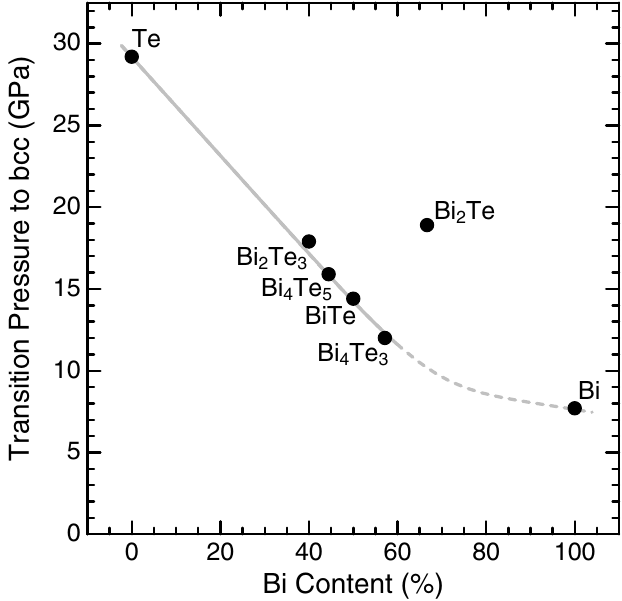}
     \caption{Transition pressures to the bcc alloy phases of \BimTen\ during compression
              as a function of composition. At the given pressures, approximately 50\% of
              each powder sample had transformed from the lower-pressure to the bcc phase.
              The transition pressures for Te and Bi are from Refs.~\onlinecite{HM03} and
              \onlinecite{AFK82}, respectively.}
     \label{fig:Bi-Te_bcc_Transition_Pressures}
\end{figure}

Figure~\ref{fig:Bi-Te_bcc_Transition_Pressures} illustrates that the pressure at which the
transition to the bcc phase occurs varies continuously across the series of compositions ---
with the notable exception of Bi$_2$Te for which the transition pressure is $\sim$9~GPa higher
than expected from the overall trend. The determination of the crystal structures that precede
bcc Bi$_2$Te at lower pressure is in progress, and the results, that may shed some light on
this anomaly, will be reported elsewhere.

\subsection{Semi-disordered B2-Type Phases}

\begin{figure}
     \centering
     \includegraphics{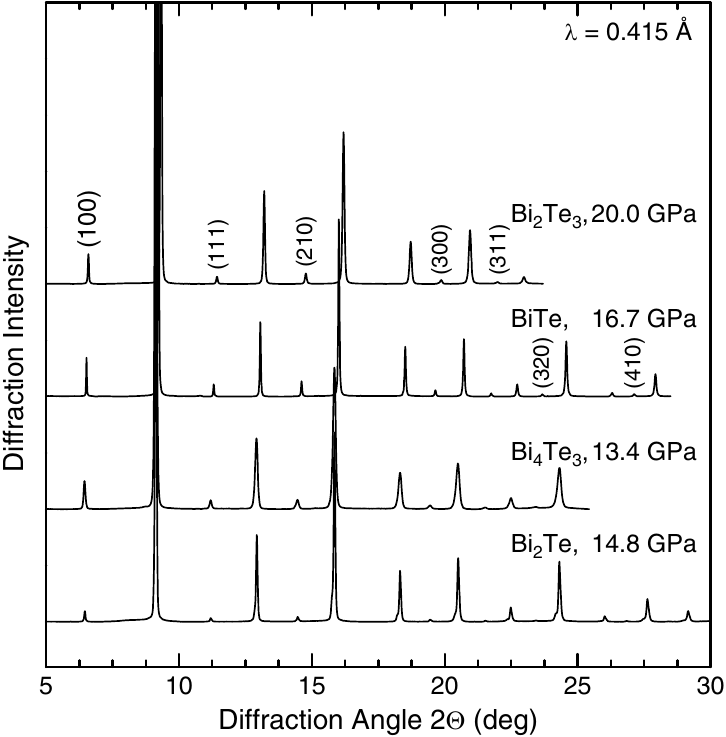}
     \caption{Powder x-ray diffraction patterns of the partially-ordered B2-type phases of
              \BimTen\ after compression at room temperature and subsequent annealing at
              100\degreeC\ (BiTe, Bi$_4$Te$_3$, Bi$_2$Te) or 120\degreeC\ (Bi$_2$Te$_3$).
              Miller indices marks the superstructure reflections.}
     \label{fig:Bi-Te_B2_patterns}
\end{figure}

In our initial experiments on Bi$_2$Te$_3$ we attempted to sharpen the diffraction pattern of
a phase at 17~GPa (phase III in the notation of Refs.~\onlinecite{NETN09,EONI11,ZWWL11}) by
annealing the sample in the DAC gently at 120\degreeC. During this treatment, the sample
transformed to a cubic phase, but not the bcc alloy phase discussed above. Rather, a
closely-related phase with superstructure reflections in the diffraction pattern was observed,
as illustrated in Fig.~\ref{fig:Bi-Te_B2_patterns}. Evidently, the thermal treatment induced
some atomic ordering, and the same effect was subsequently observed also for BiTe,
Bi$_4$Te$_3$, and Bi$_2$Te. (Bi$_5$Te$_4$ has not yet been tested for this effect.) The
diffraction patterns with the superstructure reflections can be indexed using a simple cubic
unit cell and lattice parameters similar to those of the corresponding bcc phases.

\begin{figure}
     \centering
     \includegraphics{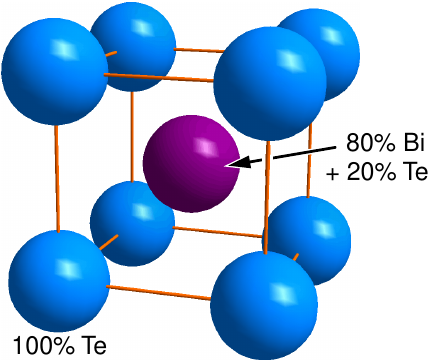}
     \caption{(Color online) Idealized crystal structure of semi-disordered B2-type
              Bi$_2$Te$_3$ with full occupation of one crystallographic site and mixed
              occupation of the other.}
     \label{fig:Bi2Te3_semiordered-structure}
\end{figure}

As it is not possible to obtain an ordered structure of Bi$_2$Te$_3$ on the basis of a cubic
unit cell containing only two atoms, we considered superstructures that are compatible with
the given stoichiometry, such as the cubic Pr$_{2}$C$_{3}$ structure type, but none of these
was consistent with the observed diffraction pattern. We also considered the monoclinic
structure {``}nine/ten-fold $C2/m${''} proposed by Zhu \etal\cite{ZWWL11}, which they identified in
their structure search calculations as the stable phase above 16~GPa. This structure model
produces a diffraction pattern similar to that observed for Bi$_2$Te$_3$ after annealing, but
with the large monoclinic unit cell containing 20 atoms, it produces also a large number of
low-intensity reflections that are not observed experimentally. Zhu \etal\ interpreted this
structure as an approximation of the disordered bcc alloy structure. Careful inspection of the
{``}nine/ten-fold $C2/m${''} structure reveals it to be actually more closely related to the B2
(CsCl) structure type, with 1/5 of the rows of Bi atoms along the cubic (110) direction being
replaced with rows of Te atoms (in an ordered fashion). Building on this realization and
taking into account the substitutional disorder before annealing, we determined the structures
of the annealed \BimTen\ phases to be semi-disordered variants of the B2 structure type. In
each case, one crystallographic site of the B2 unit cell is essentially occupied by the
majority element only, whereas the other site is randomly occupied by either element under the
constraint of the overall stoichiometry. For Bi$_2$Te$_3$, one site is then, ideally, fully
occupied by Te and the other site has a mixed occupancy by Bi and Te at a ratio of 80:20 as
illustrated in Fig.~\ref{fig:Bi2Te3_semiordered-structure}.

\begin{figure}
     \centering
     \includegraphics{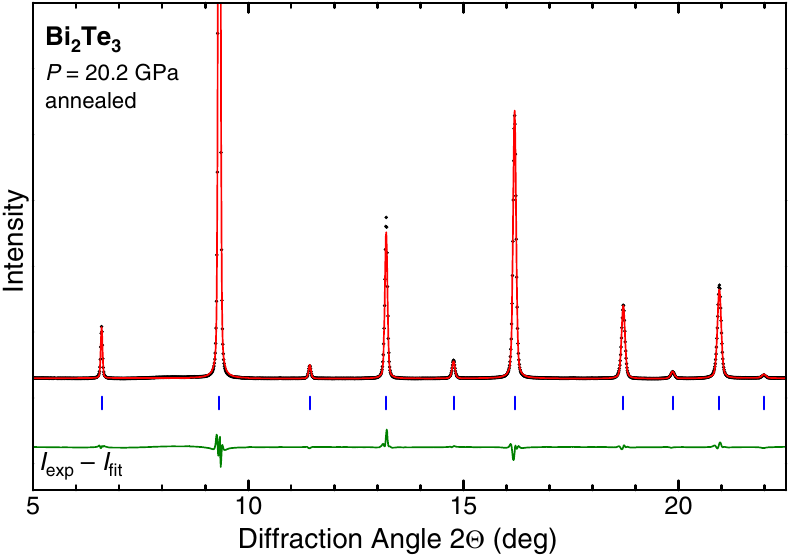}
     \caption{(Color online) Rietveld refinement for semi-disordered B2-type Bi$_2$Te$_3$
              at 20.0~GPa after 26~h of annealing at 120\degreeC. The dots  and the solid
              line in the upper part show the measured and fitted diffraction intensities,
              respectively. The tick marks show the calculated peak positions, and the
              bottom curve the difference between measured and fitted intensities.}
     \label{fig:Bi2Te3_B2_Rietveld}
\end{figure}

\begin{figure}
     \centering
     \includegraphics{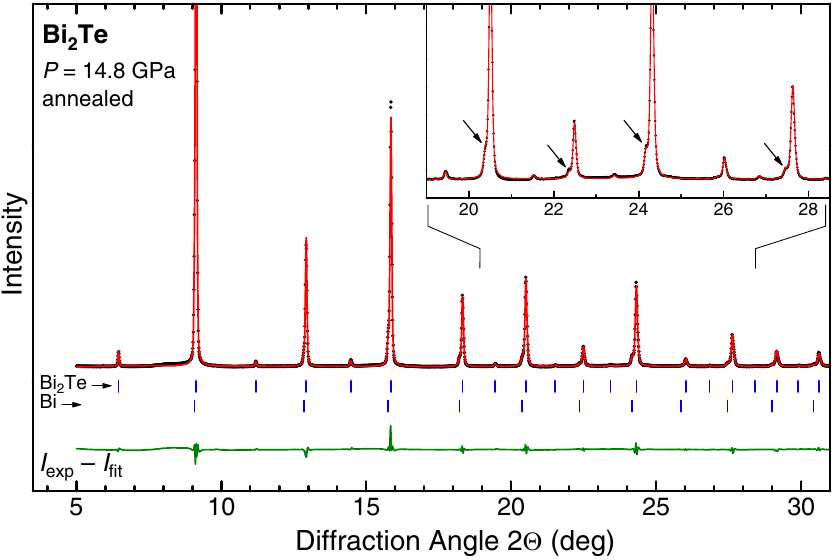}
     \caption{(Color online) Rietveld refinement for Bi$_2$Te at 14.8 GPa after 2~h of
              annealing at 100\degreeC. Two phases were refined simultaneously:
              semi-disordered B2-type Bi$_2$Te and bcc Bi. The bcc Bi phase produces
              shoulders on the lower-angle sides of some peaks as highlighted in the
              inset. The dots and the solid line in the upper part show the measured and
              fitted diffraction intensities, respectively. The tick marks show the
              calculated peak positions for the two phases, and the bottom curve the
              difference between measured and fitted intensities.}
     \label{fig:Bi2Te_B2_Rietveld}
\end{figure}

Figure~\ref{fig:Bi2Te3_B2_Rietveld} shows the result of a Rietveld refinement for the
diffraction pattern of annealed Bi$_2$Te$_3$ at 20 GPa on the basis of the semi-disordered B2
structure, which yields an excellent fit with residuals of $R_p = 1.5\%$ and $R_{wp} = 2.5\%$.
One site is mostly occupied by the majority Te with a Bi:Te ratio of 6:94(1) and the other
site consequently with a Bi:Te ratio of 74:26(1). This sample was annealed for 26~h at
120\degreeC, but we found that much shorter annealing can already induce significant ordering.
This is illustrated in Fig.~\ref{fig:Bi2Te_B2_Rietveld} which shows the result of a Rietveld
refinement for Bi$_2$Te at 14.8~GPa after only 2~h of annealing at 100\degreeC. The Rietveld
fit (with residuals of $R_p = 2.2\%$ and $R_{wp} = 3.0\%$) yields Bi:Te ratios of 90:10(2) and
41:59(2) for the two sites. This figure illustrates furthermore that we found Bi$_2$Te to
precipitate a small amount of Bi ($\sim$6\%) into a second phase. The bcc Bi phase has a
lattice parameter slightly larger\cite{AKS02} than that of B2-type Bi$_2$Te at this pressure,
so that it produces shoulders on the lower-angle sides of the Bi$_2$Te diffraction peaks as
shown in the inset to Fig.~\ref{fig:Bi2Te_B2_Rietveld}.

Table~\ref{tab:Bi-Te_annealing} summarizes the conditions under which four \BimTen\ phases
were annealed to transform them from their alloys to the semi-disordered B2 phases, together
with the relative occupancies of the two sites as a measure of the ordering. The BiTe sample
was annealed three times in succession. The first treatment at a relatively high pressure of
25~GPa (11~GPa above the transition pressure to bcc) produced only rather limited ordering.
Two subsequent annealing steps at lower pressures of 17--18~GPa eventually produced almost
complete (97\%) ordering. Altogether, we found that thermal treatment at moderate temperatures
of 100--120\degreeC\ causes the high-pressure Bi-Te bcc alloy phases to transform readily to
semi-ordered phases of the B2 structure type. Prolonged annealing of BiTe led to nearly
complete ordering, and it appears likely that for the other phases a higher degree of ordering
could also be obtained with prolonged annealing.

\begin{table}
  \centering
  \caption{Atomic ordering in B2-type \BimTen\ after thermal annealing. The temperature, duration and final
  pressure of the thermal treatment are indicated by $T$, $\Delta t$, and $P$, respectively. The Bi:Te ratios or
  relative occupancies of the two sites were determined with Rietveld refinements. Site 1 is the site of the
  majority element, and its Bi:Te ratio represents the degree of ordering. The BiTe sample was annealed three
  times in succession, with each step as indicated.}
  \medskip

    \begin{tabular}{lccc@{\extracolsep{0.5em}}D{:}{:}{4}D{:}{:}{4}}
    \hline
    \hline
    Phase & \multicolumn{3}{c}{Annealing} & \multicolumn{2}{c}{Bi:Te ratios} \\
    \cline{2-4} \cline{5-6}         & $T$ (\degreeC) & $\Delta t$ (h) & $P$ (GPa) & \multicolumn{1}{c}{site 1} & \multicolumn{1}{c}{site 2} \\
    \hline
    Bi$_2$Te$_3$ & 120 & \phantom{+}26 & 18 & 6:94(1)  & 74:26(1) \\[3pt]
    BiTe   & 100 & \phantom{+}25 & 24 & 26:74(2) & 74:26(2) \\
                & 100 & +42 & 18 & 91:9(1)  & 9:91(1)  \\
                & 100 & +92 & 17 & 97:3(1)  & 3:97(1)  \\[3pt]
    Bi$_4$Te$_3$ & 100 & \phantom{+}16 & 14 & 90:10(2) & 24:76(2) \\[3pt]
    Bi$_2$Te  & 100 & \phantom{+1}2  & 15 & 90:10(2) & 41:59(2) \\
    \hline
    \hline
    \end{tabular}%
  \label{tab:Bi-Te_annealing}%
\end{table}%

The phases with compositions other than 1:1 BiTe could potentially order further by adopting
superlattice structures of the B2 type, e.g.\ partially disordered variants of the structures
of Heusler compounds. However, careful inspection of the patterns in
Fig.~\ref{fig:Bi-Te_B2_patterns} yields no evidence for $(\half\,\half\,\half)$ superlattice
reflections that would be indicative of structure types such as L2$_1$, D0$_3$ or Y
(Ref.~\onlinecite{GFP11,KZSV15}).

\begin{figure*}
     \centering
     \includegraphics{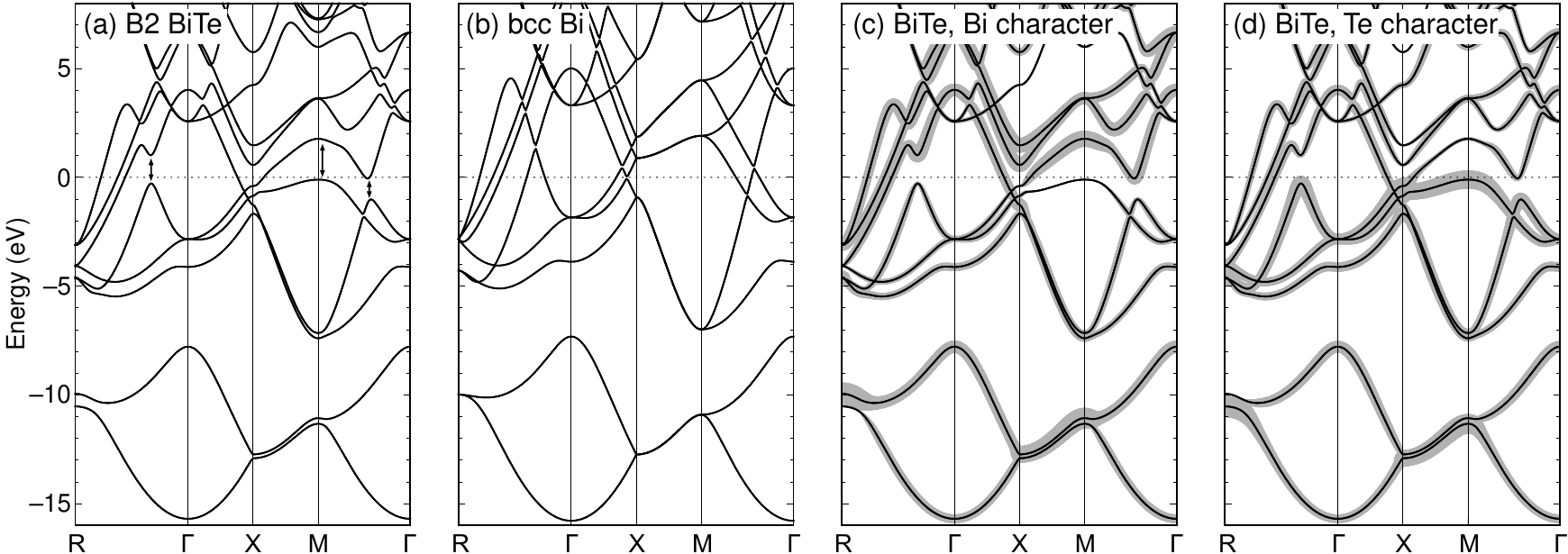}
     \caption{Calculated electronic band structures of (a) B2-type BiTe and (b) bcc Bi for
              cubic unit cells with a lattice parameter of 3.623~{\AA}, corresponding to the
              experimental density of bcc BiTe at 20~GPa. The calculated pressure for
              B2-type BiTe at this density is 18.5~GPa. To facilitate the comparison
              between B2-type BiTe and bcc Bi, the band structure for the latter is not
              shown with respect to the usual bcc Brillouin zone, but rather with respect
              to the same simple cubic Brillouin zone as for B2-type BiTe. (c)~Bi
              character and (d) Te character of the electronic bands in BiTe at 18.5~GPa
              are shown by the widths of the gray bands ({``}fat bands{''}), as quantified by
              the partial charges of the band states in the atomic spheres. The most
              notable differences between (c) and (d) can be seen for states close to the
              Fermi level, e.g.\ around the M point.}
     \label{fig:BiTe_bandstructures}
\end{figure*}

Intermetallic compounds are well known to exhibit widely varying degrees of atomic order ---
from completely disordered alloys to fully ordered compound phases --- depending on
stoichiometry, temperature, and thermal and mechanical treatment \cite{Cah95}. The
semi-disordered B2-type phases observed here are reminiscent of semi-disordered B2-type phases
observed in some X$_{2}$YZ Heusler compounds, e.g.\ Co$_{2}$FeAl at high temperature
\cite{KUKI04}, where the X atoms occupy one site of the B2 structure and the Y and Z atoms
randomly occupy the other site\cite{GFP11}. Intermetallic phases generally have a tendency to
long-range order\cite{Cah95}. To obtain an alloy, atomic ordering must be prevented, typically
by quenching a high-temperature disordered phase, or destroyed, e.g. by cold working. Although
it is therefore to be expected to observe ordering in the Bi-Te phases at elevated
temperatures, it is surprising to observe it at such relatively low temperatures of only
$\sim$100{\mbox{\textdegree}}C. It is therefore possible that atomic ordering will occur already at room
temperature on a longer time scale. In other words, the Bi-Te alloy phases may show signs of
aging.

Atomic disorder in intermetallic compounds is well documented to affect their physical
properties, e.g.\ the magnetism in Heusler compounds\cite{GFP11}, the mechanical properties of
structural alloys\cite{Cah95}, or the superconductivity in A15 intermetallics\cite{Cah95}.
Superconductivity in Bi$_2$Te$_3$ and Bi$_4$Te$_3$ has been observed in the pressure range
were their cubic alloy phases exist\cite{ZSCZ11,JLSS11}. The existence of both alloy and
semi-ordered Bi-Te phases therefore needs to be taken into account in investigations of their
physical properties, in particular in electronic structure calculations.

\subsection{Electronic Structure and Chemical Bonding in BiTe}

\begin{figure}
     \centering
     \includegraphics{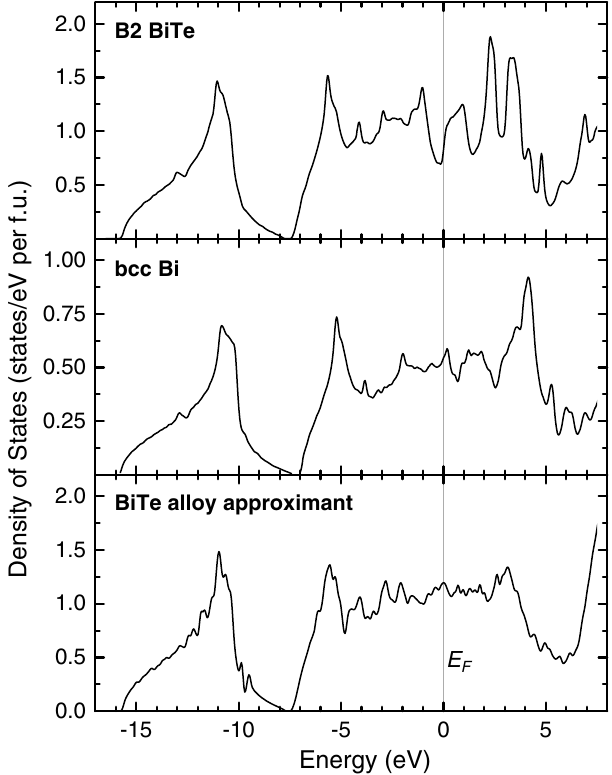}
     \caption{Calculated electronic density of states of B2-type BiTe at a calculated
              pressure of 18.5~GPa (lattice parameter $a=3.623$~{\AA}), of bcc Bi, and of a
              BiTe alloy approximant. The same atomic density was used in all three
              calculations. The Fermi level $E_F$ was set to be~0.}
     \label{fig:BiTe_DOS}
\end{figure}

The fully-ordered B2-type BiTe phase provides an ideal starting point for electronic structure
calculations, which allow us to gain insight on the nature of its chemical bonding and the
forces that drive the atomic ordering. Figure~\ref{fig:BiTe_bandstructures}(a) shows the
calculated band structure of B2-type BiTe at the experimental density of bcc BiTe at 20~GPa.
Spin-orbit coupling was included in all calculation and was found to produce significant band
splittings, in agreement with earlier studies\cite{MSJ97,TRDM92} on Bi$_2$Te$_3$. In the
calculation, BiTe is metallic at this pressure, which is in line with the observation that the
band gap in Bi$_2$Te$_3$ closes at pressures of 2--5~GPa, depending on the sample and
pressure/stress conditions.\cite{IPA64,MTZU14} The BiTe band structure is similar to that of
bcc Bi at the same atomic density [Fig.~\ref{fig:BiTe_bandstructures}(b)]. The most notable
difference is the occurrence of additional band splittings for BiTe, as indicated by arrows in
Fig.~\ref{fig:BiTe_bandstructures}(a). This reduces the number of bands crossing the Fermi
level and produces a suppression in the electronic density of states --- a pseudogap --- near
the Fermi level as illustrated in Fig.~\ref{fig:BiTe_DOS}. This suppression is absent for bcc
Bi and, more importantly, for an approximant of the BiTe alloy, which we modeled as a
$2\times2\times2$ supercell of B2-type BiTe with a random distribution of the eight Bi and
eight Te atoms. The result is a slightly reduced band structure contribution to the total
energy of the ordered phase in comparison to the alloy. This illustrates at a fundamental
level the discussion in the previous section on the effect of atomic disordering on the
electronic properties.

Analysis of the partial charges from the individual band states shows that the lower-energy
components of the split bands near the Fermi level have predominantly Te character, and the
higher-energy components predominantly Bi character [Fig.~\ref{fig:BiTe_bandstructures}(c,d)].
This indicates a charge transfer from Bi to Te, which is in keeping with Te having a slightly
higher electronegativity than Bi (2.1 vs 2.02 on the revised Pauling scale \cite{JL92}). To
quantify the charge transfer, we performed a Bader ({``}Atoms in Molecules{''})
analysis\cite{Bad91,Bad90} of the charge density distribution. In this analysis, the unit cell
is divided into regions whose boundaries are defined by minima in the charge density. More
precisely, at any point on the dividing boundaries, the gradient of the charge density is
parallel to the boundary or zero. This provides an unambiguous partitioning of the unit cell
into atomic basins, and by integrating the charge within these basins one can quantify the
charge transfer between atoms. For BiTe at $\sim$20~GPa, the Bader analysis indicates a
significant charge transfer from Bi to Te of approximately 0.44~$e$. The charge transfer
depends only weakly on the pressure and decreases with increasing pressure to 0.40~$e$ at
100~GPa.

The chemical bonding and charge transfer between Bi and Te have been analyzed in several
studies on Bi$_2$Te$_3$ at normal pressure\cite{MSJ97,Kag78,PT89,ZWWL11}. The Bi$_2$Te$_3$
crystal structure can be regarded as a repeating sequence of {``}quintuple layers{''} that comprise
five atomic layers, Te(1)--Bi--Te(2)--Bi--Te(1), where Te(1) and Te(2) denote the two
crystallographic sites occupied by Te. The bonding within the quintuple layers is of mixed
covalent-ionic type, whereas the bonding between the quintuple layers of van-der-Waals type.
The degree of ionicity is generally difficult to quantify, and model-dependent, and the
results of several studies vary correspondingly. The combined DFT and tight-binding
calculations by Mishra \etal\cite{MSJ97} gave ionic charges of $\mbox{Bi}_2^{0.7+}\,
\mbox{Te(1)}_2^{0.5-}\, \mbox{Te(2)}^{0.35-}$. An earlier tight-binding model by Pecheur and
Toussaint \cite{PT89} had indicated ionic charges about 50\% lower than those obtained by
Mishra \etal\ On the other hand, Zhu \etal \cite{ZWWL11} recently reported ionic charges of
$\mbox{Bi}_2^{0.75+}\, \mbox{Te(1)}_2^{0.05-}\, \mbox{Te(2)}^{1.39-}$ (on the basis of DFT
calculations and Bader analysis), which suggested a strong charge disproportionation between
the two Te sites. They furthermore reported the charge transfer from Bi to Te to increase with
increasing pressure. It is therefore interesting to note that the charge transfer obtained
here for cubic BiTe is significantly lower than those reported by Zhu \etal\ for the layered
ambient-pressure and three high-pressure phases of Bi$_2$Te$_3$, and their pressure
dependences are also at variance.

\begin{figure}
     \centering
     \includegraphics{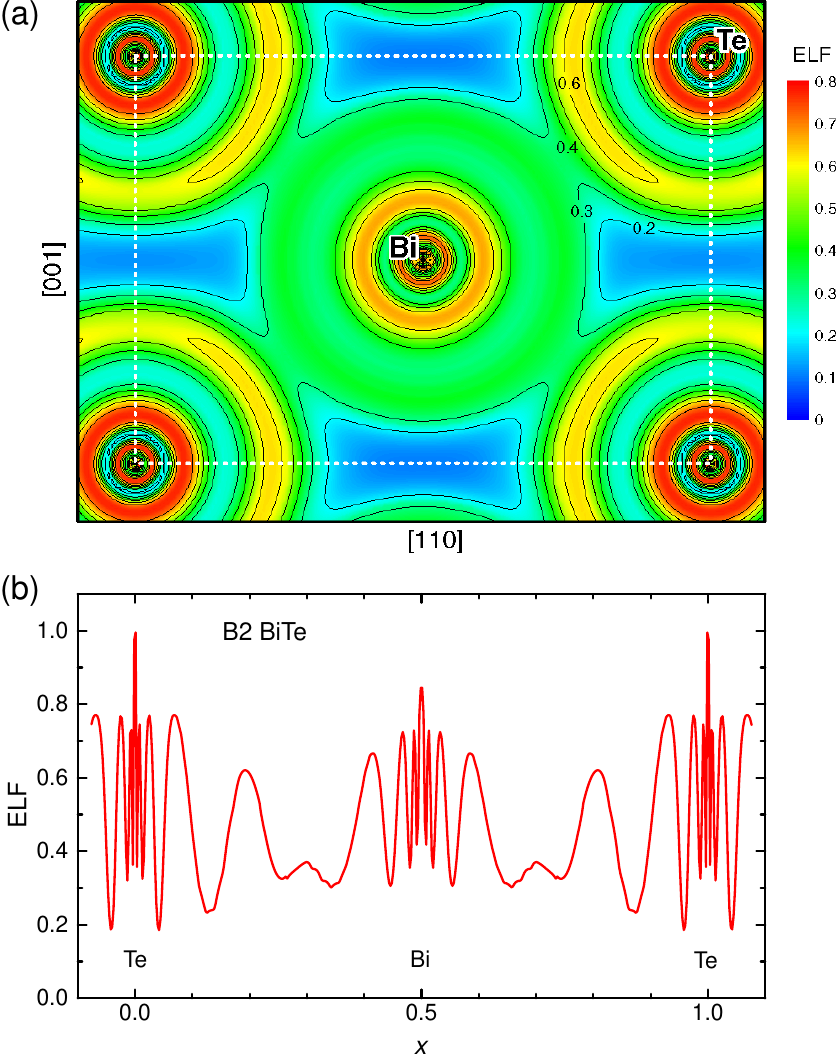}
     \caption{(Color online) Calculated electron localization function of B2-type BiTe at
              a calculated pressure of 18.5~GPa (lattice parameter $a=3.623$~{\AA}). (a)~ELF
              in the (110) plane. The unit cell edges are indicated by the white dotted
              lines. (b) ELF along the body diagonal $(x, x, x)$ of the cubic unit cell.}
     \label{fig:BiTe-ELF}
\end{figure}

To further address the question whether any significant covalent contribution to the chemical
bonding is still present at high pressure in the cubic phases (as suggested in Refs
\onlinecite{EONI11,ZWWL11}), we have calculated the electron localization function (ELF)
\cite{BE90,SNWF97} for BiTe. The results for B2-type BiTe  are shown in
Fig.~\ref{fig:BiTe-ELF}. The upper panel shows the ELF in the (110) plane that passes through
four Te atoms at the corners of the unit cell and the Bi atom in the center. There is no
indication of a local maximum of the ELF along the Bi--Te line, and hence no indication of
directional, covalent bonding between Bi and Te. The oscillations in the ELF in
Fig.~\ref{fig:BiTe-ELF}(b) represent merely the electronic shell structure of Bi and Te.

The overall picture emerging from these calculations is thus one where the mixed
covalent-ionic plus van-der-Waals bonding found in the bismuth tellurides at ambient
conditions gives way to metallic bonding with a significant ionic component at high pressures
of $\sim$20~GPa. The charge transfer from Bi to Te is rather insensitive to further increase
in pressure. We consider the charge transfer to be the driving force behind the
thermally-induced atomic ordering. Firstly, ordering of the Bi and Te ions reduces the
Madelung energy. The semi-ordered phases for the compositions other than BiTe represent the
most-ordered configurations that can be achieved within the bcc/B2 structure type under the
constraint of the given stoichiometry. Secondly, the ion ordering in BiTe causes a suppression
of the electronic density of states near the Fermi level, and hence a lowering of the band
structure contribution to the total energy. It is not clear at present to what extent this
second effect contributes to the formation of the semi-ordered phases.

\subsection{Fermi Surface of BiTe and Topological Changes}
\begin{figure}[t]
      \vspace{2ex} \centering
     \includegraphics{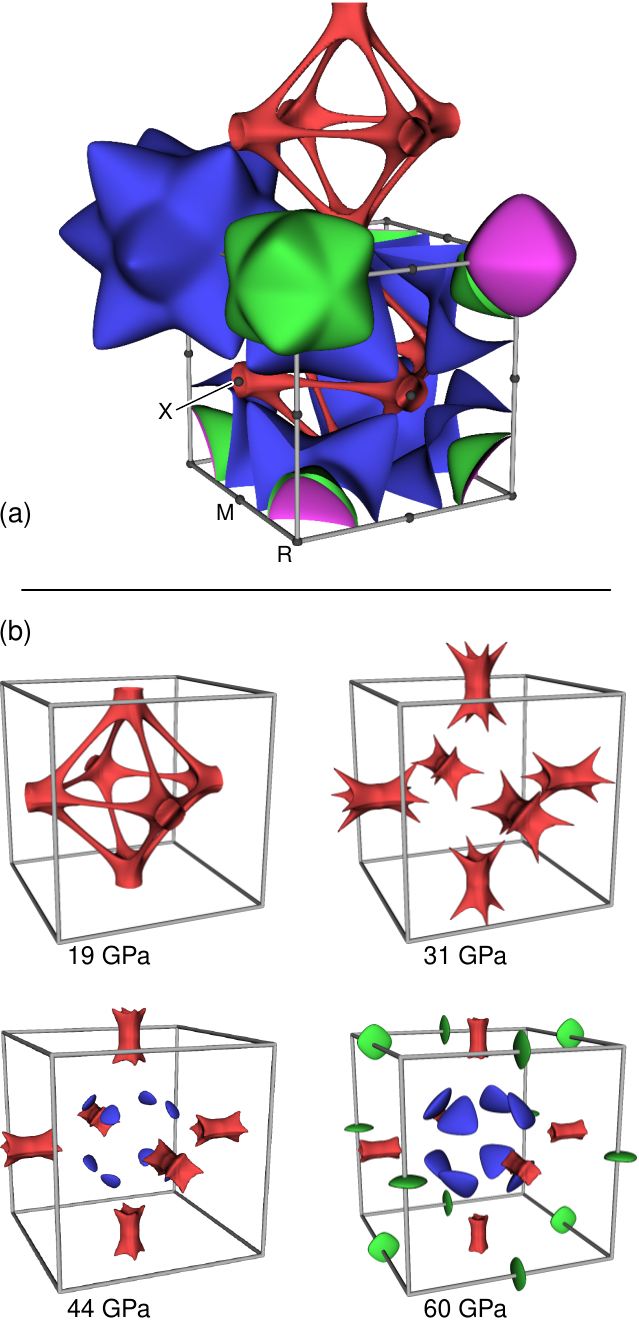}
     \caption{(Color online) (a) Fermi surface of B2-type BiTe at a calculated pressure of
              18.5~GPa (lattice parameter $a=3.623$~{\AA}). (b)~Evolution of the Fermi surface
              with pressure. The three stable pockets centered around the R point have
              been omitted for clarity.}
     \label{fig:BiTe-FS}
\end{figure}

Figure~\ref{fig:BiTe_bandstructures}(a) shows that several electronic bands cross the Fermi
level in B2-type BiTe and that there are three band extrema in close proximity to the Fermi
level. These extrema arise from the band splittings marked by arrows in
Fig.~\ref{fig:BiTe_bandstructures}(a), which are due to the ionicity in BiTe as discussed
above. The Fermi surface thus comprises several sheets, and its topology can be expected to
change with pressure when band extrema cross the Fermi level. Figure~\ref{fig:BiTe-FS}(a)
shows the calculated Fermi surface of B2-type BiTe at the same density as for the band
structure in Fig.~\ref{fig:BiTe_bandstructures}(a). It consists of three electron pockets
nested inside one another and centered around the R point as well as a tube-like feature with
necks at the X points. With increasing pressure, the band minimum along \mbox{$\Gamma$--M},
the maximum along \mbox{$\Gamma$--R}, and the maximum at M cross the Fermi level at calculated
pressures of 29, 38, and 47~GPa, respectively. The associated changes in the topology of the
Fermi surface are illustrated in Fig.~\ref{fig:BiTe-FS}(b). The tube-like feature breaks up
into pockets centered on the X points, and additional pockets appear upon further compression.
We expect these Fermi surface features to be sensitive not only to changes in pressure but
also to atomic displacements, which may give rise to interesting electron-phonon coupling
effects.

We should note that although we did not investigate the stability of cubic bcc/B2-type BiTe at
pressures above 25~GPa, we expect it to be stable over a wide pressure range.  B2-type phases
are generally rather stable under compression, and bcc Bi has been reported to be stable to at
least 222~GPa (Ref.~\onlinecite{AKS02}), bcc Te to 97~GPa (Ref.~\onlinecite{SAIF14}), and the
Bi$_2$Te$_3$ bcc alloy phase to at least 52~GPa (Ref.~\onlinecite{ZWWL11}).

\section{Conclusions}
We have shown that the formation of alloy phases at high pressure appears to be a common
feature across the series of \BimTen\ compounds. The alloy phases were found to be thermally
unstable and to adopt semi-disordered B2/CsCl-type structures after gentle annealing, but not
the superlattice structures known from Heusler compounds. The atomic ordering must be expected
to affect the physical properties of the alloy phases, e.g.\ superconductivity
\cite{II72,ETNO10,ZZWZ11,ZSCZ11,MTZU14, JLSS11, SJWV16}. Electronic structure calculations
indicate a significant charge transfer of $\sim$0.4~$e$ from Bi to Te in B2-type BiTe. The
ionic contribution to the bonding favors atomic ordering and thus the transition from alloy to
ordered compound. Another contribution is the lowering of the band structure contribution to
the total energy due to a suppression of the electronic density of states near the Fermi level
that results from the charge transfer. The calculated electron localization function for BiTe
provides no evidence for any remaining covalent bonding. The chemical bonding thus changes
from mixed covalent-ionic (plus van-der-Waals interactions between the layers)  at normal
conditions to mixed metallic-ionic at pressure of 20~GPa and above. The Fermi surface of BiTe
has an intricate structure, which may give rise to interesting electron-phonon coupling
effects, and it has been calculated to undergo three topological changes in the 20--60~GPa
range.

\emph{Note added.} After submission of the manuscript, Stillwell \etal\ \cite{SJWV16} reported
the observation of a bcc alloy phase in Bi$_2$Te under high pressure and that this phase is
superconducting. Superconductivity is certainly an interesting example of possible
electron-phonon coupling effects mentioned above. An obvious question is now how the atomic
ordering reported in the present work affects the superconducting properties of cubic \BimTen\
phases.

\acknowledgments We thank D. L. Sun and C. T. Lin (Max-Planck-Institut f\"{u}r
Festk\"{o}rperforschung, Stuttgart) for providing the Bi$_2$Te$_3$ single crystal as well as M.
Hanfland (ESRF) and A. Kleppe (DLS) for their support during the diffraction experiments.
Facilities were made available by the European Synchrotron Radiation Facility and Diamond
Light Source.

%\bibliographystyle{apsrev4-1}
%\bibliography{papers,misc,Loa_BiTeOrder}

\begin{thebibliography}{47}
\bibitem [{not()}]{note:DFT-full-details}%
  \BibitemOpen
  \href@noop {} {}\bibinfo {note} {The atomic sphere radii $R_{\text{MT}}$ were
  2.5~bohr for both Bi and Te in the Wien2k calculations, and 2.8 and 2.6~bohr,
  respectively, in the ELK calculations. The plane wave basis set was defined
  by the parameter $R_{\text{MT}} K_{\text{max}} = 9$.}\BibitemShut {Stop}%
\end{thebibliography}

%\input{Loa_BiTeOrder_160603_bbl.inc}


\end{document}